\documentclass[preprint,10pt]{article}

\usepackage{arxiv}
\usepackage{titling}
\usepackage{graphicx,color,psfrag}
\usepackage{fancyheadings}

\usepackage[backend=biber,style=alphabetic-verb]{biblatex}
\bibliography{metrics}

\newif\ifidentifying
\identifyingtrue

\title{A Simpler Method for Understanding Emergency Shelter Access Patterns}

\ifidentifying

\author{Geoffrey Messier  \\
  University of Calgary  \\
  2500 University Dr.~NW, Calgary, AB, Canada, T2N 1N4\\
  gmessier@ucalgary.ca
}

\else

\author{}

\fi

\date{}

\newcommand{\bE}{\begin{enumerate}}
\newcommand{\eE}{\end{enumerate}}
\newcommand{\bI}{\begin{itemize}}
\newcommand{\eI}{\end{itemize}}
\newcommand{\I}{\item}

\AtEveryBibitem{\clearfield{isbn}}
\AtEveryBibitem{\clearfield{issn}}

\begin{document}

\maketitle
\thispagestyle{plain}

\begin{abstract}
The Simplified Access Metric (SAM) is a new approach for characterizing emergency shelter access patterns as a measure of shelter client vulnerability.  The goal of SAM is to provide shelter operators with an intuitive way to understand access patterns that can be implemented by non-technical staff using spreadsheet operations.  Client data from a large North American shelter will be used to demonstrate that SAM produces similar results to traditional transitional, episodic and chronic client cluster analysis.  Since SAM requires less data than cluster analysis, it is also able to generate a real time picture of how shelter access patterns are affected by external factors.  Timelines generated from nine years of shelter client data using SAM demonstrate the impact of Housing First programming and the COVID-19 lockdown on how people access shelter.  Finally, SAM allows shelter staff to move beyond assigning transitional, episodic and chronic labels and instead use the ``soft'' output of SAM directly as a measure of vulnerability.
\end{abstract}

\keywords{emergency shelter, housing first, vulnerability indicators, individual trajectories, chronic homelessness, COVID-19 pandemic}

\section{Introduction}
\label{sec:Intro}

Different people access shelter in different ways.  Identifying different types of shelter access patterns has become an established way of better understanding homelessness.  Starting in 1998, cluster analysis has been used to shed light on three different patterns of shelter access \cite{kuhnApplyingClusterAnalysis1998}.  Temporary or transitional clients access shelter for a very short time.  Chronic clients make very regular use of shelter over a long period.  Episodic clients also access shelter over a very long period but on a much more irregular basis.  Since this first study, clustering has been used repeatedly to understand and compare patterns of homelessness in many different contexts \cite{kneeboneWhoAreHomeless2015, aubryIdentifyingPatternsEmergency2013, mcallisterRethinkingResearchForming2011, culhaneTestingTypologyFamily2007}.  

Understanding shelter access patterns is not just a research question.  It is also an important part of delivering housing and support services in practice.  People experiencing chronic homelessness have been identified as a high priority group for Housing First services \cite{namianGoverningHomelessnessInstruments2020}.  However, shelter operators and housing program agencies currently do not have a good way to identify chronic shelter access.  Many definitions exist for chronic homelessness \cite{byrneTestingAlternativeDefinitions2015} that could be applied to a client's shelter access record.  However, these require a client to spend a long time in shelter which is obviously contrary to the Housing First philosophy.  Cluster techniques require information technology (IT) infrastructure and data analysis expertise that many not-for-profit groups do not have.  Clustering is also restricted to imposing a ``label'' on clients and is typically performed with long records of client data.

Due to these challenges, most shelters and housing programs resort to using vulnerability indices (VI's) as a means to prioritize clients for services \cite{namianGoverningHomelessnessInstruments2020}.  However, VI's tend to lump many different risk factors into a single ``score''.   This considerable oversimplification has led even the developers of the vulnerability index - service prioritization decision assistance tool (VI-SPDAT), arguably the most prominent VI, to call for the discontinuation of its use \cite{dejongMessageOrgCodeVISPDAT2022}.  Instead, communities are being encouraged to consider multiple measures of vulnerability or risk when allocating housing and other support resources \cite{shinnAllocatingHomelessServices2022}.  It is important that shelter and housing program staff have a way to include a person's pattern of accessing emergency shelter services as one of these measures of vulnerability.  

The primary contribution of this paper is to propose a simple and practical method for summarizing a person's pattern of shelter access that meets the following criteria: 

\bI

\I {\bf Accessible:}  Any method needs to work with software and IT expertise already commonplace within shelters and housing providers.  It also must be implemented and understood by staff with non-technical backgrounds.

\I {\bf Fast:}  Any indicator of vulnerability used to allocate supports needs to yield results as quickly as possible after a client first enters shelter.

\I {\bf Empowering:} A measure of vulnerability should nuanced and allow for interpretation by human decision makers rather than imposing a hard label like ``chronic'' or ``transitional'' on large groups of clients.

\I {\bf Accurate:} While meeting all of the above criteria, this new approach needs to still generate results that are comparable to established cluster techniques in the literature.  

\eI

The proposed method for summarizing client shelter access patterns is called the Simplified Access Metric (SAM).  SAM is accessible since it can be implemented using commonly avaialable spreadsheet programs.  An open source spreadsheet is provided \cite{} in association with this paper that allows shelter providers to test SAM on their own client data \cite{messierSAMDemonstrationSpreadsheet2022}.  Unlike cluster techniques that operate using a client's entire historical record of shelter access, SAM is faster.  It yields accurate results that are comparable to clustering while using only the first few months of a client's shelter record.  While SAM can be used to assign the traditional ``chronic'', ``episodic'' and ``transitional'' labels, its output can be used without labels to have a more nuanced understanding of how a client accesses shelter.  This empowers staff to combine SAM with other vulnerability indicators to make better resource allocation decisions.   

A second contribution of this paper is to use SAM on a large shelter client data set to illustrate the how the introduction of Housing First programming and the COVID-19 pandemic have affected client shelter access patterns over time.  Homelessness is a dynamic phenomenon and there is considerable benefit to understanding how it changes over time due to a variety of factors, as discussed in  \cite{leeHomelessnessMovingTarget2021} and associated articles.  Since SAM is able to produce results with very little client data, it can provide snapshots of the state of client shelter access at any particular time.  When many of these snapshots are viewed over time, trends can be observed.  

SAM will be evaluated using over 12 years of client data collected at the Calgary Drop-In Centre (DI), a large North American emergency homeless shelter in Calgary, Canada.  The accuracy of SAM will be established by comparing it to established clustering techniques.  The paper will then demonstrate how SAM goes beyond clustering by using it to reveal dynamic trends in shelter access and discussing how it can be used without assigning labels to clients.

\section{Methods}
\label{sec:Methods}

\subsection{Data Set}
\label{ssec:Data}

This is a secondary data analysis performed on anonymized shelter stay records collected at the Calgary Drop-In Centre (DI) between July 1, 2009 and March 31, 2022.  The data anonymization and client privacy protocol used during this study was approved by the University of Calgary Conjoint Faculties Research Ethics Board (REB19-0095).  A waiver of consent was received for this study since it deals with anonymized secondary data.  The data set consists of 1,904,044 entries collected for 24,512 clients.  An entry consists of an anonymized client identification number and a date.  Each entry represents a single client accessing shelter sleep services one or more times during the 24 hour period indicated by the entry date stamp.  While the analysis in this paper is limited to data from a single shelter, it has been shown in \cite{messierBestThresholdsRapid2022} that the data is representative of several other shelters in North America.

External factors can have a significant impact on how people access shelter services.  The more than 12 years represented in the DI data set spans the introduction of Housing First programming at the DI and the COVID-19 pandemic.  To isolate and investigate the effect of these external factors, the data is divided into three eras.  The {\em Housing Ready} era includes data from the beginning of the data set to August 1, 2017 which is the approximate beginning of Housing First programming at the DI.  The {\em Housing First} era includes entries from August 1, 2017 to March 1, 2020 which marks the approximate beginning of the COVID-19 pandemic in Calgary.  The {\em COVID-19} era includes the entries from March 1, 2020 to the end of the data set.

\subsection{The Simple Access Metric Framework}
\label{ssec:Sam}

SAM characterizes a client's shelter access pattern using:

\bI
\I {\bf Shelter Duration}: The number of days since a client first accessed shelter.
\I {\bf In-Shelter Percentage}: The percentage of days during a client's shelter duration that they accessed shelter.
\I {\bf Active Client Threshold}:  A client is considered ``active'' if the number of days since their last shelter access is less than this threshold.  For example, if the active client threshold is 30 days, all clients who have accessed shelter within the last 30 days would be considered active.
\eI

Consider a client who has accessed shelter on April 10, 11, 14 and 22.  Today is April 30 and the active client threshold is 10 days.  This client would be considered active since their last access was 8 days ago.  Their shelter duration would be 20 days since their first access was April 10.  Since they have accessed shelter 4 times over 20 days, their in-shelter percentage would be 20\%.

There is a tradeoff when deciding when to calculate a client's active status, shelter duration and in-shelter percentage.  If done too soon after a client first enters shelter, not enough shelter entries will be accumulated to show a clear pattern. If done too long after first entering shelter, connecting the client to support services may be delayed.  While each shelter is free to determine a time that suits their program cycles and client population, we will demonstrate in Section~\ref{ssec:Cohorts} that 90 days after first entering shelter works well for the DI client data set.

SAM can be mapped onto the concepts of transitional, chronic and episodic access patterns in an intuitive way.  Transitional clients exit shelter very quickly and would most likely fall into the inactive client category when evaluated.  Chronic shelter clients are regular shelter users who would tend to meet the active client criterion and have both long shelter durations and high in-shelter percentages.  Episodic clients also accumulate long shelter durations but their in-shelter percentages would be lower due to the gaps in their access patterns.  

The gaps in episodic access patterns also present a tradeoff when choosing the active client threshold.  The threshold should be large enough to allow most episodic clients to show up as active even with the gaps that can be present in their access patterns.  However, an active threshold that is too long will start to include too many short term transitional clients and possibly overwhelm staff resources.  Section~\ref{ssec:Cohorts} will demonstrate that an active client threshold of 30 days produces good results for the DI data set.

\section{Results}
\label{sec:Results}

\subsection{Creating Shelter Access Pattern Cohorts}
\label{ssec:Cohorts}

As discussed in Section~\ref{sec:Intro}, cluster analysis works by assigning labels to clients.  Section~\ref{sec:Discussion} will elaborate on how SAM does not necessarily have to assign labels.  However, in order to compare SAM to cluster analysis, this section will demonstrate how SAM can be used to identify transitional, episodic and chronic clients.  This comparison will be conducted using client data from the Housing Ready era in the DI data described in Section~\ref{ssec:Data}.  In order to select a group who accessed shelter under relatively similar conditions, only clients whose entire shelter records fit inside the Housing Ready era are selected.  A total 13,295 (54.2\% of the 24,512 overall client population) meet this criterion.

After a client reaches a shelter duration of 90 days, SAM classifies a client as follows:
\bI
\I All inactive clients are transitional, where the active client threshold is 30 days.
\I All active clients with in-shelter percentages greater than the threshold $\alpha$ are chronic.
\I All active clients with in-shelter percentages less than or equal to the threshold $\alpha$ are episodic.
\eI

Assigning a final label to a client after 90 days is primarily to compare SAM to clustering.  In a real shelter, a client who is transitional after 90 days according to the above rules may re-enter shelter on the 91st day and proceed to access shelter on a very regular basis.  Of course, staff will continue to interact with that client and evaluate them for services.   SAM is completely compatible with this approach.  Staff would simply continue to update that client's shelter duration, in-shelter percentage and active client status as described in Section~\ref{ssec:Sam}.  The rules above could be used to relabel the client at a later date or, as we recommend in Section~\ref{sec:Discussion}, the staff person could abandon labeling entirely and evaluate the client's shelter access pattern using shelter duration and in-shelter percentage directly for as long as they remain active.

To generate the clustering results, a K-means cluster algorithm ($K = 3$) is applied to the total number of stays and total number of shelter stay episodes for each client \cite{kneeboneWhoAreHomeless2015, aubryIdentifyingPatternsEmergency2013, mcallisterRethinkingResearchForming2011, culhaneTestingTypologyFamily2007}.  A stay episode is defined as a series of shelter stays with gaps between stays of less than 30 days.  

Since the objective is to prove that SAM produces similar results to cluster analysis, the transitional/episodic/chronic classifications produced by cluster analysis are treated as ``true''.  Then, the accuracy of SAM can be defined as the number of clients that receive the same label from the two algorithms divided by the total number of clients.  The $\alpha$ threshold was swept from 5\% to 95\% in increments of 5\% to find the value of $\alpha$ that gives the best accuracy.  The best result is achieved using $\alpha$ = 85\% which yields an accuracy of 81.6\%.

While numerical performance metrics like accuracy can be useful, the real objective of both cluster analysis and SAM is to identify cohorts of people who access shelter in a similar way.  To this end, Table~\ref{tb.AccStat} presents the statistics for the total number of stays and episodes for clients classified using cluster analysis and SAM.  For example, Table~\ref{tb.AccStat} indicates that cluster analysis labels 219 clients as chronic (1.65\% of the total cohort of 13,295 clients).  This group stays in shelter a mean of 1018.11 times, a median of 843 times and the upper 10th percentile of this group has accumulated 1782 or more stays.

\begin{table}[htbp]
\centering
\begin{tabular}{r|ccc}
                         &                          &          Stays           &         Episodes        \\
                         &            N             & (mean/median/10th pctl)  & (mean/median/10th pctl) \\
\hline
  Transitional - cluster &  11984/13295 (90.14\%)   &      20.62 / 2 / 53      &       1.52 / 1 / 3      \\
      Transitional - SAM &  11007/13295 (82.79\%)   &      16.29 / 2 / 28      &       1.78 / 1 / 3      \\
      Episodic - cluster &   1092/13295 (8.21\%)    &    115.34 / 65 / 311     &      7.01 / 6 / 10      \\
          Episodic - SAM &   1668/13295 (12.55\%)   &    102.40 / 49 / 255     &       3.36 / 3 / 7      \\
       Chronic - cluster &    219/13295 (1.65\%)    &   1018.11 / 843 / 1782   &       2.50 / 2 / 5      \\
           Chronic - SAM &    620/13295 (4.66\%)    &    396.75 / 218 / 947    &       2.02 / 1 / 4      \\
\end{tabular}
\caption{Client shelter access statistics.}
\label{tb.AccStat}
\end{table}

\subsection{Shelter Access Patterns Over Time}
\label{ssec:Timeline}

\begin{figure}[htbp]
\centerline{\includegraphics[width=4in]{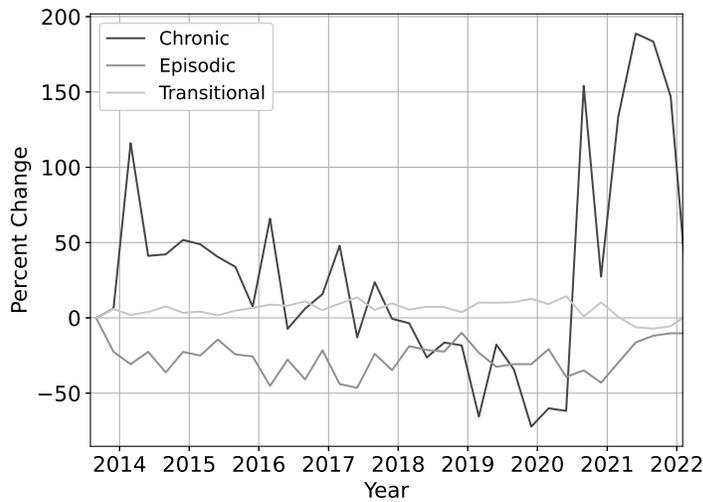}}
\caption{Percent change in shelter access pattern groups.}
\label{fg.PctChange}
\end{figure}

This section illustrates how SAM can be used to investigate variations in shelter access patterns over time.  SAM is used as described in Section~\ref{ssec:Cohorts} to classify the shelter access patterns of each client who accessed shelter from mid-2013 onward.  The number of transitional, chronic and episodic clients identified by SAM are then summed up in three month intervals.  Figure~\ref{fg.PctChange} illustrates the percent change in these three month totals relative to the start of the timeline.  

\begin{figure}[htbp]
\centerline{\includegraphics[width=4in]{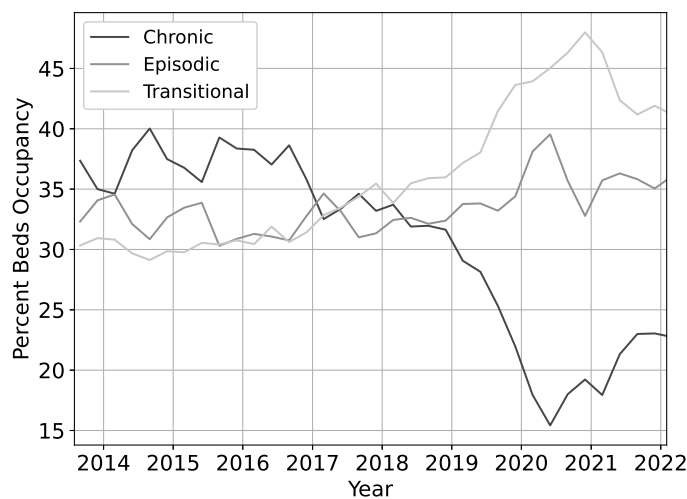}}
\caption{Percent bed occupancy for shelter access pattern groups.}
\label{fg.SysUse}
\end{figure}

Fig.~\ref{fg.PctChange} shows the largest relative variation in the chronic shelter access and it perhaps is tempting to dismiss this variation as not significant since clients with chronic access patterns make up a small percentage of the total client population.  However, it is well known in the homelessness serving sector that this small group still utilizes a large proportion of shelter resources.  To demonstrate this, Fig.~\ref{fg.SysUse} shows what percentage of total bed occupancy in each three month interval are due to clients in the transitional, episodic and chronic shelter access groups.

Note that cluster analysis could be used on historical data to create plots that are similar to Figs.~\ref{fg.PctChange} and \ref{fg.SysUse}.  Clustering could be performed on the historical data records of all the clients in the DI data set.  The 3 month totals shown in the figures could then be calculated using the cluster assigned labels for each client and their date of first entry to shelter.  However, unlike clustering, it is important to emphasize that SAM can generate Figs.~\ref{fg.PctChange} and \ref{fg.SysUse} in real time.  For example, since SAM uses only 90 days of client data, it can generate the 3 month totals in Fig.~\ref{fg.PctChange} at the start of 2019 using only data from that 3 month period.  Since clustering is typically performed on a client's entire historical data record, the labels assigned to clients who first enter shelter at the start of 2019 may require data from months or even years after the clients' first entry date.  This means that clustering could generate an academically interesting retrospective analysis on historical client data but it is not useful for assessing client vulnerability in real time.

\section{Discussion}
\label{sec:Discussion}

SAM meets the objectives laid out in Section~\ref{sec:Intro}.  Length and percentage of time spent in shelter are intuitive since they map to concepts like ``they have been here a long time'' and ``they are here all the time''.  The open source spreadsheet \cite{messierSAMDemonstrationSpreadsheet2022} provided with this paper demonstrates that these quantities are easily calculated using date, ranking and filter spreadsheet operations.  Given the simplicity of these metrics, they produce results surprisingly comparable to the much more complex clustering method.  Selecting cohorts consisting of 81.6\% of the same clients overall is strong performance, especially considering SAM makes its selections using only 90 days of client data.  

That said, it is a mistake to focus too much on numerical performance metrics since the purpose of both cluster analysis and SAM is to divide clients into groups with similar characteristics.  When considering Table~\ref{tb.AccStat}, it is tempting to point out that there is considerable relative error between SAM and clustering in some cases.  For example, clustering identifies a group of chronic clients with a mean of 1018.11 stays while the SAM chronic group has a mean of 396.75 stays.  The number of shelter episodes for the episodic groups is also lower for SAM.  The reduced ability of SAM to discriminate access patterns is the cost of its simplicity.  However, the performance of SAM is still sufficient for use as a measure of client vulnerability.  For example, it can certainly be argued that any group of people with an average of 396.75 shelter stays is accessing shelter on a very regular basis and will likely not exit shelter without being connected to supports.

Section~\ref{ssec:Timeline} demonstrates that SAM can provide some interesting insights into how external factors affect the way people access shelter.  The impact of Housing First programming and the COVID-19 pandemic at the DI was significant for the group with chronic access patterns in particular.  Fig.~\ref{fg.PctChange} shows that Housing First programming significantly reduced the size of the chronic group from its introduction mid 2017 to the start of the pandemic.  The proportion of clients with chronic access patterns then jumped considerably during the pandemic.  Anecdotally, DI staff indicated that this was largely due to the ``shelter in place'' directives being communicated during lockdown.  A reduction in the chronic access patterns is seen in 2022 as shelter operations return to normal but more data is required to see if this drop will be sustained.  Fig.~\ref{fg.SysUse} shows people with chronic access patterns consuming the most shelter resources until a dramatic reduction with the introduction of Housing First.  This climbs again during the pandemic but not enough to erase the gains achieved by Housing First.

Finally, while much of this paper has been dedicated to the discussion of the transitional, episodic and chronic labels, it is the author's opinion that these labels should be abandoned.  Any method that assigns a label must use some kind of hard threshold.  A person who just misses meeting that threshold may have almost identical characteristics to another who receives the label.  In fact, the unlabeled person may be an even better candidate for supports if other measures of vulnerability are taken into account, as recommended in \cite{shinnAllocatingHomelessServices2022}.  Instead of a label, providing a human decision maker with a ``soft'' measure that summarizes a client's shelter access history allows them to better balance the vulnerability suggested by shelter access history with other factors.  SAM can easily be used label free.  A shelter or housing staff member can simply look directly at a client's shelter duration and in-shelter percentage figures to gain an understanding of their shelter access history.  Cluster analysis provides no such option.

\section{Conclusion}
\label{sec:Concl}

The Simplified Access Method (SAM) technique presented in this paper is a way for shelter and housing program staff to characterize client shelter access patterns as one measure of client vulnerability.  SAM is entirely based on the concepts of percentage of time in shelter and total length of shelter interaction.  This means it is both intuitive and easy to implement using commonplace spreadsheet tools.  Similar to cluster analysis, SAM can label a client's shelter interaction as transitional, episodic and chronic.  Even though SAM uses a much smaller window of client data than clustering, it produces similar results.  However, the real benefit of SAM is that it allows shelter staff to dispense with labeling clients entirely and to instead use shelter duration and in-shelter percentage values directly as an indicator of vulnerability.

\section{Acknowledgments}
\label{sec:ack}

The authors would like to acknowledge the support of the Natural Sciences and Engineering Research Council of Canada (NSERC), the Calgary Drop-In Centre and the Government of Alberta.  This study is based in part on data provided by Alberta Seniors, Community and Social Services. The interpretation and conclusions contained herein are those of the researchers and do not necessarily represent the views of the Government of Alberta. Neither the Government of Alberta nor Alberta Seniors, Community and Social Services express any opinion in relation to this study.

\printbibliography

\end{document}